\documentclass[useAMS,usenatbib,babel]{mn2e}

\usepackage[english,english]{babel}
\usepackage{amsmath}
\usepackage{amssymb,amsfonts,textcomp}
\usepackage{array}
\usepackage{supertabular}
\usepackage{hhline}
\usepackage{hyperref}
\usepackage[usenames]{color}
\hypersetup{dvips, colorlinks=true, linkcolor=blue, citecolor=blue, filecolor=blue, urlcolor=blue}
\usepackage[dvips]{graphicx}

\def\gtrsim{\lower.5ex\hbox{$\; \buildrel > \over \sim \;$}}
\usepackage{graphicx}

\newcommand{\hagn}{\mbox{{\sc \small Horizon-AGN\,\,}}}

\definecolor{grey}{rgb}{0.75,0.75,0.75}
\definecolor{Orange}{rgb}{1.0,0.5,0.15}
\definecolor{brown}{rgb}{0.7,0.25,0.0}
\definecolor{pink}{rgb}{1.0,0.5,0.5}
\definecolor{darkerred}{rgb}{0.8,0,0}
\definecolor{darkerblue}{rgb}{0,0,0.8}
\definecolor{Blue}{rgb}{0,0.08,0.65}
\definecolor{Red}{rgb}{0.65,0.08,0.05}
\definecolor{Green}{rgb}{0.15,0.45,0.25}

\begin{document}

\author[C. Welker,  et al.]{
%\parbox[t]{\textwidth}{Y. Dubois$^{1,2}$\thanks{E-mail: dubois@iap.fr} et al}
\parbox[t]{\textwidth}{C. Welker$^{1,2}$,  J.~Devriendt$^{2}$,  Y. Dubois$^{1,2}$, C. Pichon$^{1}$ and
S. Peirani$^{1}$}
\vspace*{6pt} \\
$^{1}$ Institut d'Astrophysique de Paris, UMR 7095, CNRS, UPMC Univ. Paris VI, 98 bis boulevard Arago, 75014 Paris, France.\\
$^{2}$ Sub-department of Astrophysics, University of Oxford, Keble Road, Oxford OX1 3RH, United Kingdom.\\
%$^{4}$ Observatoire de Lyon, UMR 5574, 9 avenue Charles Andr\'e, Saint Genis Laval 69561, France\\
}
\date{Accepted . Received ; in original form }

\title[Mergers  drive spin swings along the cosmic web]{
Mergers  drive spin swings along the cosmic web 
% On the importance of spin swings induced by galaxy mergers
}

\maketitle

\begin{abstract}
{The close relationship between mergers and the reorientation of the {\it spin}  for galaxies and their host dark haloes is investigated using  a cosmological hydrodynamical simulation (\hagn\!\!). Through a statistical analysis of merger trees, we show that  spin swings are mainly driven by mergers along the filamentary structure of the cosmic web, and that these events account for the preferred perpendicular orientation of massive galaxies with respect to their nearest filament. By contrast, low-mass galaxies ($M_{\rm s}<10^{10} \,Ê\rm M_\odot$ at redshift 1.5) undergoing very few mergers, if at all, tend to possess a spin well aligned with their filament. Haloes follow the same trend as galaxies but display a greater sensitivity to smooth anisotropic accretion. The relative effect of mergers on spin magnitude is qualitatively different for minor and major mergers: mergers (and diffuse accretion) generally increase the magnitude of the angular momentum, but the most massive major mergers also give rise to a population of objects with less spin left. Without mergers secular accretion builds up the spin of galaxies but not that of haloes. It also (re)aligns galaxies with their filament.
}
\end{abstract}

\begin{keywords}
%cosmology: theory ---
%galaxies: evolution ---
galaxies: formation ---
galaxies: haloes ---
galaxies: kinematics and dynamics ---
large-scale structure of Universe ---
methods: numerical
\end{keywords}

%===============================================
% Introduction
%===============================================
\section{Introduction}

Over the past ten years, several numerical investigations~\citep[e.g.][]{calvoetal07, hahnetal07b, pazetal08, sousbie08} have reported that large-scale structures, i.e. cosmic filaments and sheets, influence the direction of the   angular momentum (AM) of haloes, in a way originally predicted by~\cite{lee&pen00}. 
It has been speculated that massive haloes have AM perpendicular to the filament and higher spin parameters because they are the results of major mergers~\citep{aubertetal04,peirani2004,bailin&steinmetz05}.
On the other hand, low-mass haloes acquire most of their mass through smooth accretion, which explains why their AM is preferentially parallel to their closest large-scale filament~\citep{codisetal12, laigle2014}. 
Using the cosmological hydrodynamical \hagn simulation, \cite{duboisetal14} have shown that this trend extends to galaxies: the AM of low-mass, rotation-dominated, blue, star-forming galaxies is 
preferentially aligned with their  filaments, whereas high-mass, velocity dispersion-supported, red quiescent galaxies tend to possess an AM perpendicular to these filaments. 
These theoretical predictions have recently received their first observational support~\citep{Tempel13}.  Analysing Sloan Digital Sky Survey (SDSS) data, these authors uncovered a trend for spiral galaxies to align with 
nearby structures, as well as a trend for elliptical galaxies to be perpendicular to them.

In this letter, we revisit these significant findings with an emphasis both on exploring the physical mechanisms which drive halo's and galactic spin swings and on quantifying how much mergers and smooth accretion re-orient these spins relative to cosmic filaments.
After a brief review of the numerical methods in Section~\ref{section:virtual}, we analyse the effect of mergers and smooth accretion on spin orientation and magnitude for haloes and galaxies in Section~\ref{section:galaxy}.

%=============================================== 
%  setup
%===============================================
\section{Numerical method}
\label{section:virtual}

The cosmological hydrodynamical simulation analysed in this paper, \hagn\!\!, is already described in~\cite{duboisetal14}, so we only summarize its main features in this letter.
We adopt a standard $\Lambda$CDM cosmology with total matter density $\Omega_{\rm m}=0.272$, dark energy density $\Omega_\Lambda=0.728$, amplitude of the matter power spectrum $\sigma_8=0.81$, baryon density $\Omega_{\rm b}=0.045$, Hubble constant $H_0=70.4 \, \rm km\,s^{-1}\,Mpc^{-1}$, and $n_s=0.967$ compatible with the WMAP-7 data~\citep{komatsuetal11}.
The size of the simulated volume is $L_{\rm box}=100 \, h^{-1}\rm\,Mpc$ on a side, and it contains $1024^3$ dark matter (DM) particles, which results in a DM mass resolution of $M_{\rm DM, res}=8\times 10^7 \, \rm M_\odot$.
The simulation is run with the {\sc ramses} code~\citep{teyssier02}, and the initially uniform grid is adaptively refined down to $\Delta x=1$ proper kpc at all times. 
Refinement is triggered in a quasi-Lagrangian manner: if the number of DM particles becomes greater than 8, or the total baryonic mass reaches 8 times the initial DM mass resolution in a cell.
Gas can radiatively cool down to $10^4\, \rm K$ through H and He collisions with a contribution from metals using rates tabulated by~\cite{sutherland&dopita93}. 
Heating from a uniform UV background takes place after redshift $z_{\rm reion} = 10$ following~\cite{haardt&madau96}. 
The star formation process is modelled using a Schmidt law: $\dot \rho_*= \epsilon_* {\rho / t_{\rm ff}}$ for gas number density above $n_0=0.1\, \rm H\, cm^{-3}$,  where $\dot \rho_*$ is the star formation rate density, $\epsilon_*=0.02$ the constant star formation efficiency, and $t_{\rm ff}$ the local free-fall time of the gas.
Feedback from stellar winds, supernovae type Ia and type II are also taken into account for mass, energy and metal release.
Black holes (BH) formation is also included, and they accrete gas at a Bondi-capped-at-Eddington rate and coalesce when they form a tight enough binary.
BHs release energy in a quasar (heating) or radio (jet) mode when the accretion rate is above (below) one per cent of Eddington, with efficiencies tuned to match the BH-galaxy scaling relations~\citep[see][for details]{duboisetal12agnmodel}.

Galaxies and haloes are identified with the AdaptaHOP finder~\citep{aubertetal04} which operates on the distribution of star and DM particles respectively with the same parameters than in~\cite{duboisetal14}.
Unless specified otherwise, only structures with a minimum of $N_{\rm min}=100$ particles are considered, which typically selects objects with masses larger than  $2\times10^8 \, \rm M_\odot$ for galaxies and $8\times10^9 \, \rm M_\odot$ for DM haloes. Catalogues containing up to $\sim 150 \, 000$ galaxies and $\sim 300 \, 000$ DM haloes are produced for each redshift output analysed in this letter ($1.2 < z < 3.8$).
The spin of a galaxy is defined as the total AM of the star particles it contains and is measured with respect to the densest of these star particles (centre of the galaxy).

The galaxy (halo) catalogues are then used as  an input to build merger trees with TreeMaker~\citep{tweedetal09}. 
Any galaxy (halo) at redshift $z_n$ is connected to its progenitors at redshift $z_{n-1}$ and its child at redshift $z_{n+1}$. 
We build merger trees for 18 outputs from $z=1.2$ to $z=3.8$ equally spaced in redshift.
On average, the redshift difference between outputs corresponds to a time difference of 200 Myr (range between 100 and 300 Myr).
We reconstruct the merger history of each galaxy (halo) starting from the lowest redshift $z$ and identifying the most massive progenitor at each time step as {\it the galaxy} or {\it main progenitor}, 
and the other progenitors as {\it satellites}. Moreover, we double check that the mass of any child contains at least half the mass of its main progenitor to prevent misidentifications.
Note that the definition of mergers (vs smooth accretion) depends on the threshold used to identify objects as any object composed of fewer particles is discarded and considered as smooth accretion.
Finally, in order to get rid of objects too contaminated by grid-locking effects (grid/spin alignment trend for the smallest structures, see~\citealp{duboisetal14}), we exclude galaxies with $M_{\rm s} <10^9 \, \rm M_\odot$ and haloes with $M_{\rm h} <10^{11} \, \rm M_\odot$ from our {\it main progenitor} sample for spin analysis.
Satellites, however, can be smaller structures, which is why we adopt a low object identification mass threshold, and select more massive {\it main progenitors} afterwards.
This two-step procedure allows for a clear separation of main progenitors and satellites and avoids significant signal loss.
Note that, in all figures where haloes and galaxies are compared, the ratio of main progenitor minimal mass to satellite minimal mass is the same, so as to permit a fair comparison between both categories of objects.

In order to quantify the orientation of galaxies (haloes) relative to the cosmic web, we use a geometric three-dimensional ridge extractor called the ``skeleton''~\citep{sousbie09} computed from a density cube of $512^3$ cells drawn from the simulation and gaussian-smoothed with a smoothing length of $3 \, h^{-1}\, \rm Mpc$ comoving. The orientation of the spin of galaxies (haloes) can then be measured relative to the direction of the closest filament segment.

%=============================================== 
%  Galactic spin evolution via mergers 
%=============================================== 
\section{Spin swings and mergers}
\label{section:galaxy}

 %--------------------------------------------------------------------------------------
\begin{figure}
\center \includegraphics[width=0.7\columnwidth]{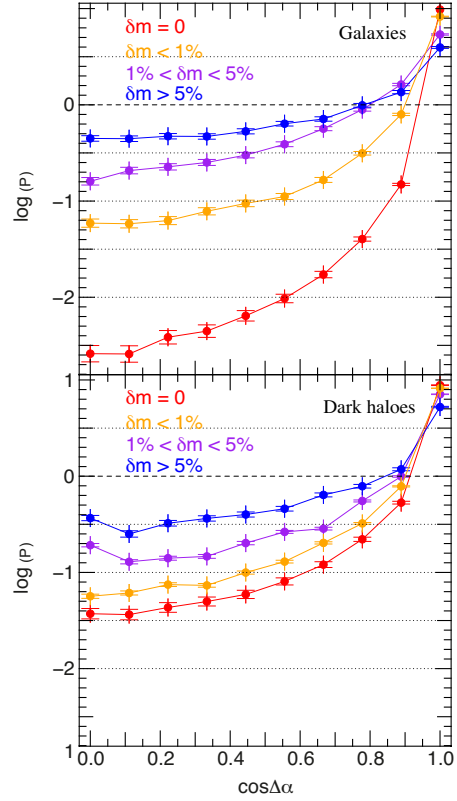}
  \caption{Logarithm of the PDF of $\cos\Delta\alpha$,  the cosine of the spin swing angle for galaxies (top panel) and haloes (bottom panel) between two time steps, for objects with different merger histories. The dashed line corresponds to the uniform PDF, i.e.  no preferred orientation. The dotted lines show the threshold below which the ratio per bin falls under 30\%, 10\%, 3\% and 1\% of the $\delta m$-sample considered. $\delta m$ is the mass fraction accreted through mergers  between  two consecutive time outputs. $\delta m=0$ corresponds to the no merger case, i.e. pure smooth accretion. Mergers are responsible for spin swings; haloes are more sensitive to smooth accretion.}
\label{fig:dm_flip}
\end{figure}
%--------------------------------------------------------------------------------------
 
First, we define $\delta m=\Delta m_{\rm mer}(z_n)/M(z_n)$ as the mass fraction of an object that is accreted via mergers. In this expression, $M(z_n)$ is the total stellar (DM) mass of a galaxy (halo) at redshift $z_n$ and 
$\Delta m_{\rm mer}(z_n)$ is the stellar (DM) mass accreted by this galaxy (halo) through mergers between redshifts $z_{n-1}$ and $ z_n$. In a similar spirit, we also define the relative AM variation of an object between two 
consecutive simulation outputs as $\delta \lambda= (L_{n}- L_{n-1})/(L_{n}+ L_{n-1})$, where $L_{n}$ is the magnitude of the object AM at redshift $z_{n}$. 

Fig.~\ref{fig:dm_flip} (top panel) displays the Probability Distribution Function (PDF) of $\cos\Delta\alpha$, where $\Delta \alpha$ is the variation in the angle of the galaxy spin between two consecutive time outputs, for galaxies with different merger histories, i.e. different values of $\delta m$. We recall that the satellite detection threshold is set at $N_{\rm min}=100$ particles, but that only main progenitors with masses $M_{\rm s} > 10^9 \, \rm M_\odot$ (galaxies) and 
$M_{\rm h} > 10^{11}\, \rm M_\odot$ (haloes) are considered. 
From this figure, one can see that mergers are clearly the main drivers for galaxy spin swings, while the spins of galaxies without mergers tend to remain aligned between time outputs. Indeed, $93\%$ of these latter see their spin stay within an angle of $25 \deg$ between two consecutive time outputs ($\Delta z= 0.1$) whereas this happens only for $40\%$ of galaxies with a merger mass fraction above $5\%$ (this ratio even falls down to $15\%$ with $N_{\rm min}=1000$). 
Such a swing effect is sensitive to the merger mass fraction and, as one would expect, tends to be stronger for larger fractions. For $\delta m>5\%$, $35\%$ of the galaxy sample underwent a spin swing $> 50 \deg$ while this is true for only $16\%$ of galaxies with $1\%<\delta m<5\%$ and $0.2\%$ of the no-merger ($\delta m = 0$) population.
However, even mergers with low mass ratio (i.e. mergers where the satellite is less than twenty times lighter than the main progenitor) trigger important swings compared to the no-merger case. Only $79\%$ of the galaxies which underwent a minor merger ($\delta m<1\%$) maintain a spin within a cone of $25 \deg$ between consecutive time outputs (compared to $93\%$ for non-mergers). This behaviour is consistent with the well-known fact that when two galaxies merge, the remnant galaxy acquires a significant fraction of AM through the conversion of the orbital AM of the pair rather than simply inheriting the intrinsic spin of its progenitors. 

A similar analysis for DM haloes confirms that they qualitatively follow the same behaviour as galaxies but with quantitative variations due to the fact they are velocity dispersion-supported structures rather than rotationally supported ones. 
More specifically, one can see from Fig.~\ref{fig:dm_flip} that unlike galaxies, even haloes defined as non-mergers ($\delta m = 0$) exhibit noticeable spin swings~\citep[see also][]{bettetal2012}. This can be attributed to the net AM of haloes resulting from random motions of DM particles (by opposition to ordered rotational motion of star particles 
for galaxies): even a small amount of AM brought in coherently by smooth accretion or mergers will be enough to noticeably influence the direction of the halo spin vector. Note that large-scale tidal torques also apply more efficiently to haloes than galaxies due to the larger spatial extent of the former, and we speculate that these torques could also contribute to some of the quantitative differences we measure between AM alignment of haloes and galaxies.

%
%--------------------------------------------------------------------------------------
\begin{figure}
\center\includegraphics[width=0.80\columnwidth]{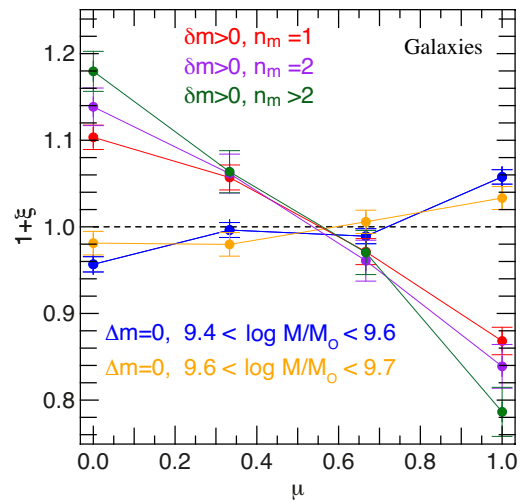}
  \caption{PDF of $\mu$, the cosine of the angle between the galactic spin and its  filament  for different galaxy merger histories. This plot shows cumulative results for all simulation galaxies identified between $z=3.16$ and $z=1.71$. $\xi$ is the excess probability with respect to a uniform distribution (dashed line). As before, $\delta m$ is the fraction of mass accreted through mergers between two consecutive time outputs, and $n_m$ is the total number of mergers a galaxy has undergone at the time of the measurement. $\delta m=0$ corresponds to the absence of mergers. The stronger the merger rate the stronger the misalignment. Subsequent mergers amplify the alignment.
}
\label{fig:gal_align}
\end{figure}
%--------------------------------------------------------------------------------------
%
Given that mergers account for the spin swings of galaxies,  they should  also be responsible for setting the orientation of their spins relative to the filament, at least for massive galaxies which do experience a significant amount of mergers. 
Our results are  consistent with this scenario, as can be seen in Fig.~\ref{fig:gal_align} where we plot the PDF of $\mu$, the cosine of the angle between the galactic spin and the direction of its filament, $\xi$ being the excess probability
with respect to a uniform distribution. It demonstrates that galaxies (each one being counted once after each merger) which have just merged tend to be more perpendicular to filaments, and that the signal is stronger for galaxies which have experienced a larger number of mergers 
during their lifetime. This is a strong argument in favour of orbital AM transfer into spin since mergers are preferentially the result of galaxies encounters along cosmic filaments, i.e., pairs with an orbital AM that is orthogonal to the filament.
Note that the excess probability $\xi\simeq 0.1-0.2$  of being perpendicular to their filament for galaxies undergoing mergers is larger than when the same galaxies are simply split in sub-samples according to their physical properties: mass, colour, activity, etc. ($\xi <0.05$ in that case, see~\citealp{duboisetal14}).
 
In contrast, galaxies with no merger are more likely to be aligned with their filament. Note that the threshold for structure detection here  was set to $N_{\rm min}=1000$ particles, which implies that ``merger'' galaxies are more clearly identified than ``non-merger'' ones in this figure. The alignment signal is therefore weaker, as expected. To emphasise this selection effect, the excess probability of alignment was analysed for galaxies split in different mass bins, the  lowest two  of which we plot in Fig.~\ref{fig:gal_align}. Comparing both measurements, there is indeed tentative evidence that the excess probability of alignment is weaker for higher mass galaxies, which are more likely to have accreted ``undetected'' mergers. Note that the alignement signal is completely lost when we consider that sub-sample of galaxies with masses above $10^{10}\, \rm M_\odot$.
Further analysis confirms that lower thresholds ($N_{\rm min}<1000$) attenuate the orthogonal misalignment and strengthen the alignment excess probabilities.

\begin{figure}
\center \includegraphics[width=0.72\columnwidth]{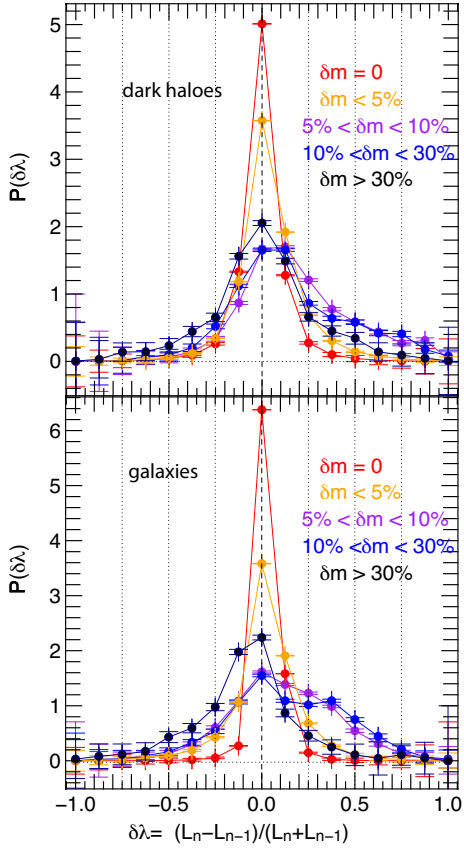}
  \caption{PDF of $\delta \lambda$ of  the halo spins (top panel) and galaxy spins (bottom panel), for objects with different merger ratios. The horizontal dashed line corresponds to the uniform PDF, i.e. the predicted distribution if the merger keeps no memory of the pre-merger spin. Positive values correspond to objects which acquire AM through mergers, negative values correspond to objects which lose AM. This plot shows results for the entire population 
of objects identified between $z=3.8$ and $z=1.2$. 
Mergers increase  the AM's magnitude, major mergers ($\delta m>0.3$) notwithstanding.
 }
\label{fig:dm_normL}
\end{figure}
%--------------------------------------------------------------------------------------

Turning to the magnitude of the AM, Fig.~\ref{fig:dm_normL} shows the PDF of $\delta \lambda$ for both galaxies and haloes. We can see from this figure that mergers with mass ratios $10\%<\delta m < 30\%$ tend to increase the magnitude of the object spin (curves are skewed towards positive $\delta \lambda$),  and that this effect becomes stronger as the mass ratio increases, up to mass fractions around $10<\delta m < 30\%$ for which $\sim$75\%  of haloes and galaxies see their spin magnitude increase -- by a factor 2 or more for $\sim$25\% of haloes and galaxies -- between two consecutive time outputs. For higher mass ratios, the measured trend is not as clear: major mergers between haloes tend to randomise (symmetric shape of the PDF) rather than increase the AM magnitude. A  decreasing trend is even measured for the AM magnitude of galaxies (curve skewed towards negative $\delta \lambda$), yielding a population of child galaxies with a lower amount of rotation in this case. 

\begin{figure}
\center \includegraphics[width=0.75\columnwidth]{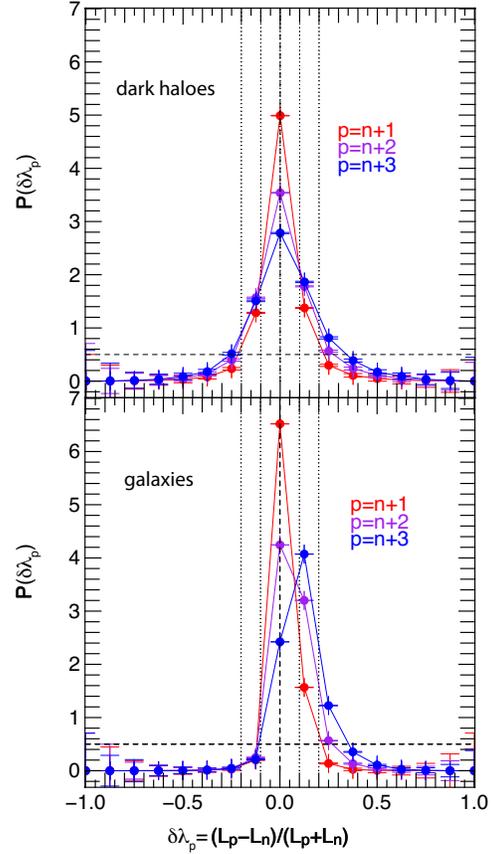}
  \caption{Same as Fig.~\protect\ref{fig:dm_normL} for objects which do not merge and for different lookback times.
   Secular accretion builds up the spin of galaxies but not that of haloes.}
\label{fig:dm0_normL}
\end{figure}

This behaviour indicates that most mergers contribute constructively to the AM of the collapsed structures. This is especially true for halo mergers where it can be understood as the conversion of orbital AM into intrinsic AM of the massive host. For minor ($\delta m < 5\%$) to intermediate ($ 5\%< \delta m < 10\%$) galaxy mergers, satellites are most likely progressively stripped of their gas and stars and swallowed in the rotation plane of the central object, therefore increasing this later rotational energy. However, major mergers  ($\delta m > 10\%$) -- where an important part of the rotation energy can be converted to random motion energy through violent relaxation, intense star formation and feedback -- can in fact contribute destructively to the AM of the galaxy remnant. Major mergers between haloes induce wings in the PDF of $\delta \lambda$ corresponding to  haloes with  increasing and decreasing  AM.
	
With $\delta m=0$ the PDF bends towards positive $\delta \lambda$, suggesting that smooth gas accretion on galaxies, unlike smooth DM accretion on haloes tends to increase their AM over time. 
In order to probe this (re)alignment process further, we present in Fig.~\ref{fig:dm0_normL} the evolution of the PDF of $\delta \lambda_p \equiv (L_{p}- L_{n})/(L_{p}+ L_{n})$, where $L_{p}$ is the AM magnitude at redshift $z_{p}$ and $p =n+1, n+2, n+3$ indicates different lookback time outputs, for haloes and galaxies. 
It appears clearly that while the halo distribution remains symmetric over time, the galaxy distribution shifts towards positive values with an average peak drift timescale of $t_{\delta \lambda}\simeq5-10\,Ê\rm Gyr$.
We measure a similar trend for different galaxy mass bins up to $M_{\rm s}=10^{11}\, \rm M_\odot$ (albeit with a slower drift for the most massive galaxies with $M_{\rm s} \approx 10^{11}\, \rm M_\odot$). 

\begin{figure}
\center \includegraphics[width=0.70\columnwidth]{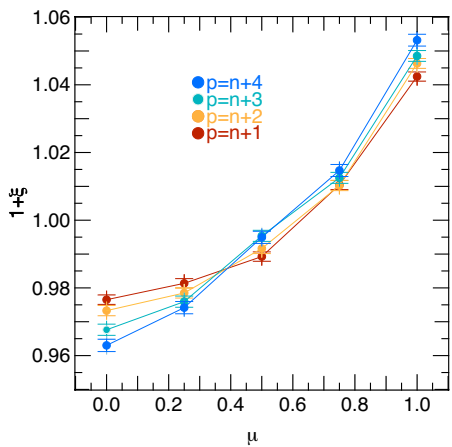}
  \caption{Same as Fig.~\ref{fig:gal_align} for galaxies which do not merge and for different lookback times.
  In absence of merger, galaxies tend to re-align with their filament over time.}
\label{fig:dm0_align}
\end{figure}

These findings strongly favour the idea that anisotropic streams of cold gas spin up galaxies over time.
This secular gas accretion onto galaxies also (re)aligns the galaxy with its filament.  This is demonstrated in
Fig.~\ref{fig:dm0_align}, which is obtained via stacking for four successive time steps the relative orientation of the spins of galaxies to filaments 
when no merger occurs.  It shows that the excess probability of alignment  is  amplified with time in the absence of mergers.

%--------------------------------------------------------------------------------------
%--------------------------------------------------------------------------------------

To sum up, \cite{Tempel13}, found that spiral galaxies tend to have a spin aligned to their nearest filament while the spin of S0 galaxies are more likely to show an orthogonal orientation. 
\cite{duboisetal14} argue that a transition mass can be associated to this change in spin orientation, which is reasonably bracketed between $\log(M_{\rm s}/\rm M_\odot)=10.25$ and $\log(M_{\rm s}/\rm M_\odot)=10.75$.
These authors also point out that such a mass loosely corresponds to the characteristic mass at which a halo extent becomes comparable  to that of the vorticity quadrant in which it is embedded within its host filament~\citep{laigle2014}. 
Such a mass dependent scenario was first suggested by~\cite{hahnetal07b}, and  quantified by~\cite{codisetal12} for DM haloes. The key idea which underpins all these studies is that lighter galaxies  acquire most of their spin through secondary infall  from their (aligned with the filament) vorticity rich environment,  while more massive galaxies acquire a large fraction of theirs via orbital momentum transfer during merger events which mainly take 
place along the direction of the large scale filament closest to them.
This letter showed that galaxies without merger both realign to their host filament and increase their AM, while successive mergers 
drive the  remnant's spin perpendicular to it, and depending on the strength of the merger, decrease  or redistribute the remnant's spin magnitude.
Hence it strongly favours the idea that cold flows feed low-mass disc galaxies with anisotropic gas streams (along the vorticity rich filaments), enhancing their AM magnitude over time, as advocated in \cite{pichonetal11}. It also demonstrates that mergers are  responsible for the 
spin swings as suggested by previous investigations.

%==============================================
% Conclusion
%==============================================
\section{Conclusion}
\label{section:conclusion}
Using the \hagn cosmological gas dynamics simulation we have analysed the variations of spin orientation and magnitude of galaxies and haloes as a function of their merger rates.
Our statistical analysis of merger trees, shows that spin swings are driven by mergers, which have a strong impact on both the {\sl orientation} and the {\sl magnitude} of the spin.
Our findings are the following:
\begin{itemize}
\item the stronger the strength of the merger the  larger the memory loss of  the post-merger spin direction of dark haloes and galaxies;
\item the alignment of the spin of an object with the cosmic web depends on its merger history: the more mergers contribute to its mass, the more likely its spin will be perpendicular to its filament;
\item when the merger contribution to the mass of an object is negligible ($ < 1\%$) the modulus of the spin of {\sl galaxies} still increases with time via smooth accretion.
 Moreover, the orientation of these spins drift towards (re-)alignment with the filament; this does not happen to the spin of dark matter haloes, whose magnitude remains independent of time on average; 
\item mergers (with mass ratios 0.1 to 0.3), like smooth accretion, also tend to build up the spin modulus but of both haloes and galaxies in this case; most massive major mergers (mass ratios 0.3 to 1), on the other hand, produce a low spin tail in the magnitude distribution.
\end{itemize} 

\vskip 0.1cm
%==============================================
% Acknowledgments
%==============================================
%\section*{Acknowledgments}
{\sl This work was granted access to the HPC resources of CINES (Jade) under the allocation 2013047012 made by GENCI.
%This research is part of the Spin(e) and Horizon-UK projects.
%Let us thank D.~Munro for freely distributing his {\sc \small  Yorick} programming language and opengl interface (available at {\tt http://yorick.sourceforge.net/}).
%We  warmly thank S.~Rouberol for running  the {\tt Horizon} cluster on which the simulation was  post-processed. 
This work is partially supported by the grant  ANR-13-BS05-0002. Special  thanks S.~Rouberol and S.~Codis.
}
\vspace{-0.5cm}
\bibliographystyle{mn2e}
\bibliography{author}

\end{document}